\documentclass[aps,prb,twocolumn,showpacs,superscriptaddress,groupedaddress]{revtex4-1}
\usepackage{hyperref}
\usepackage{sidecap}
\usepackage{amsmath}
\usepackage{tabularx}
\usepackage{bm}
\usepackage{euscript}
\usepackage{graphicx}
\usepackage{color}
\usepackage{amsfonts}
\usepackage{exscale}
\usepackage{amsbsy}
\usepackage{subfigure}
\usepackage{textcomp}
\usepackage{comment}

\def\ex{\eta_x}
\def\ey{\eta_y}

\begin{document}

\title{Effect of strain inhomogeneity on a chiral p-wave superconductor}
\author{Yue Yu$^1$ and S. Raghu,$^{1,2}$}
\affiliation{$^{1}$Department of physics, Stanford University, Stanford, California 94305, USA}
\affiliation{$^2$SLAC National Accelerator Laboratory, 2575 Sand Hill Road, Menlo Park, CA 94025, USA}
\date{\today}

\begin{abstract}
Motivated by recent measurements of strain effects on the transition temperature (T$_c$) of Sr$_2$RuO$_4$, we study the strain response of a two-dimensional chiral p-wave supercoductor.  We focus on  the effects of inhomgeneous strain fields,  which are always present in any such experiment, and which have been neglected in previous theoretical treatments.  We show that the response of T$_c$ of a chiral superconductor to strain chages from being linear  without inhomogeneity, to quadratic in the presence of inhomogeneity.  We discuss our results in the context of the ongoing debate of superconductivity in Sr$_2$RuO$_4$.  
\end{abstract}

\maketitle

\section{Introduction} 
Strontium ruthenate (Sr$_2$RuO$_4$)  is in many ways an archetypal unconventional superconductor\cite{Mackenzie2003,Mackenzie2017,Kallin2012}.   The layered perovskite can be grown to a high degree of perfection.  Normal state properties of the material are known to  unprecented detail\cite{Bergemann2003}.   Importantly, superconductivity in this material develops  out of a Fermi liquid normal state.  One might therefore expect that the superconducting properties of this system can eventually be explained from   well-controlled theories of unconventional pairing.  

Until recently, the consensus based on experimental observations, was that the superconducting ground state of Sr$_2$RuO$_4$ has chiral $p_x+ip_y$ symmetry.  Such a superconductor has odd parity, and spontaneously breaks time-reversal symmetry, since the order parameter belongs to a doubly degenerate irreducible representation of the tetragonal point group.  A consequence of such a superconducting state, is that the superconducting transition must split when the tetragonal symmetry is explicitly broken, say by an in-plane magnetic field, or by the application of uniaxial stress.  Gauge invariance requires that such perturbations couple at lowest order to the modulus squared of the order parameter; consequently, T$_c$, the superconducting transition temperature, is predicted to vary linearly with  strain.  

However, neither an in-plane magnetic field\cite{Mao2000,Yaguchi2002} or uniaxial strain\cite{Hicks2014} result in a linear variation of T$_c$.  These findings have cast some doubt on whether Strontium ruthenate is a chiral p-wave superconductor.  Motivated by the more recent strain measurements, we revisit the question of uniaxial strain effects on two dimensional chiral p-wave superconductors.  In particular, we focus on the effect of strain inhomogeneity, which are always present in any real experiment (see, for instance Ref. \onlinecite{Watson2018}, which provides quantitative estimates of such strain inhomogeneity).  Our key result is simply that local variations in strain qualitatively alters the behavior of T$_c$, causing it to vary quadratically as a function of weak average strain.  

The paper is organized as follows.  In section \ref{gl}, using Ginzburg-Landau theory, we briefly review the effect of homogeneous strain on a chiral p-wave superconductor and present the phase diagram in the presence of strain inhomogeneity.    Section \ref{discussion} contains a brief discussion ofthe results within the context of other experiments in Strontium Ruthenate.   

\section{Ginzburg-Landau description of a  $p_x + i p_y$ superconductor with strain}
\label{gl}
Consider a two dimensional tetragonal system in the absence of quenched disorder.  In the presence of spin-orbit coupling, there is a single odd-party superconducting state, which spontaneously breaks time-reversal: the so-called chiral $p_x+ip_y$ state, with order parameter
\begin{equation}
\Delta_{\sigma \sigma'}(\bm k) =  i \vec d(\bm k) \cdot  \left( \vec \sigma \sigma^y \right)_{\sigma \sigma'} \cdot ,  \ \ \ \vec d(\bm k) \propto \left( k_x \pm i  k_y \right) \hat z,
\end{equation}
where $\hat z$ is normal to the plane in which the superconductor lives, and the $\pm$ correspond to the two possible ground states.  Time-reversal symmetry is spontaneously broken, and one ground state is chosen over the other.  The Ginzburg-Landau (GL) free energy of such a superconductor, valid in the vicinity of the phase transition, is a function of a two-component order parameter, $\left( \eta_x, \eta_y \right)$, each of which is a complex function of position and time; they  are related  respectively to the $k_x$ and $k_y$ components  of the condensate.  
 For simplicity, we neglect the Cooper pair spin degree of freedom as it will not play an essential role in what follows.   The GL free energy takes the form 
 \begin{equation}
\begin{split}
&f[\eta]=f_2[\eta]+f_4[\eta]+f_{grad}[\eta]\\
&f_2[\eta]=(a+\varepsilon+\delta{\varepsilon})|\ex|^2+(a-\varepsilon-\delta{\varepsilon})|\ey|^2\\
&f_4[\eta]=b_1(|\ex|^2+|\ey|^2)^2+\frac{b_2}{2}(\ex^{*2}\ey^2+c.c.)+b_3|\ex|^2|\ey|^2\\
&a = a_0(T-T^{(0)}_c)\\
&f_{grad}[\eta]=K_1(|\partial_x\ex|^2+|\partial_y\ey|^2)+K_2(|\partial_y\ex|^2+|\partial_x\ey|^2)\\&+[K_3(\partial_x\ex)^*(\partial_y\ey)+K_4(\partial_y\ex)^*(\partial_x\ey)+c.c.]\\&
\label{e0}
\end{split}
\end{equation}
The quantity $\varepsilon$ is the average uniaxial strain taken for simplicity to be in the B$_{1g} ( i.e., ``x^2 - y^2")$ channel, and $T_c^{(0)}$ is the mean field critical temperature in the absence of strain. We allow for local variation  of strain $\delta{\varepsilon}$, which is taken to be a quenched random variable sampled from a gaussian distribution having zero mean and variance $\sigma$:
\begin{equation}
 \overline{\delta{\varepsilon}(r)}=0, \ \ \overline{\delta{\varepsilon}(r)\delta{\varepsilon}(r')}=\sigma^2\delta(r-r'),
 \end{equation}
 where the overline denotes averaging with respect to disorder.  

The quartic couplings determine whether the chiral p-wave state at zero strain is favored over a time-reversal invariant triplet superconductor.  In what follows, we set $b_2>0$ and $4b_1-b_2 + b_3 > 0$, which ensures that the chiral p-wave state, $\eta_y  = \pm i \eta_x$ is the favored broken symmetry state for $T< T_c^{0}$  (see, for instance Ref. \onlinecite{Fischer2016}). 

\subsection{$\sigma=0$}
\begin{figure}[htb]
\centering
\includegraphics[width=5cm]{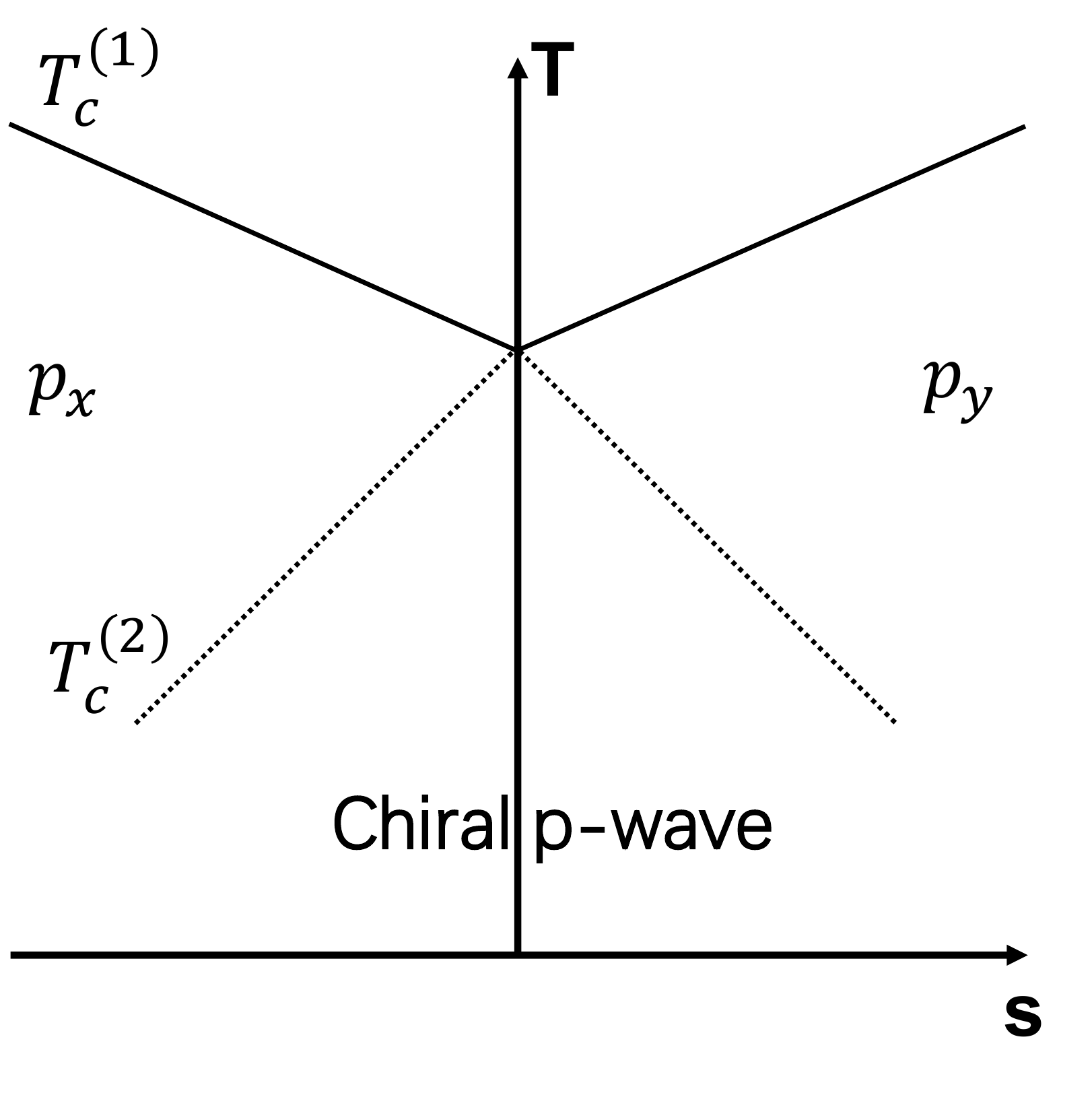}
\caption{Mean-field phase diagram in the absence of disorder. The condition $b_2>0$ and $4b_1-b_2 + b_3 > 0$ is taken such that chiral p-wave state is preferred for $T<T_c^{(2)}$. Linear phase boundaries are obtained by minimizing the effective energy density Eq. \ref{e0}. The mean field solution predicts a kink in the phase boundary at $\varepsilon=0$.}
\label{f1} 
\end{figure}

We  begin by reviewing the phase diagram in the presence of uniform strain ($\sigma=0$).   In mean-field theory, when the  gradient terms in Eq. \ref{e0} are neglected, the prediction is that strain splits the superconducting transition into two transitions: a normal-superconductor transition at temperature $T_c^{(1)}$ , and a second transition, $T_c^{(2)}$ where time-reversal symmetry is spontaneously broken: 
\begin{equation}
\begin{split}
&T_c^{(1)}=T_c^{(0)}+|\frac{\varepsilon}{a_0}|\\
&T_c^{(2)}=T_c^{(0)}-|\frac{\varepsilon}{a_0}|\frac{4b_1-b_2+b_3}{b_2-b_3}
\end{split}
\end{equation}
The corresponding mean field phase diagram is shown in Fig.\ref{f1} (see also Ref. \onlinecite{Fischer2016}). The phase boundary between the normal  and superconducting states 
is linear in $|\varepsilon|$, and has a discontinuous first derivative at $\varepsilon=0$.

For completeness,
we list the mean field solution of $|\ex|^2$ and $|\ey|^2$ as a function of the average strain $\varepsilon$ and temperature $T$
\begin{equation}
\begin{split}
&T>T_c^{(1)}:|\ex|^2=|\ey|^2=0\\
&p_x\text{ phase}:|\ex|^2=\frac{-\varepsilon-a(T)}{2b_1},|\ey|^2=0\\
&p_y\text{ phase}:|\ey|^2=\frac{\varepsilon-a(T)}{2b_1},|\ex|^2=0\\
&T<T_c^{(2)}:\ex\pm{i}\ey\text{ phase},|\ex|^2=\frac{-a(T)}{4b_1-b_2+b_3}-\frac{\varepsilon}{b_2-b_3},\\&|\ey|^2=\frac{-a(T)}{4b_1-b_2+b_3}+\frac{\varepsilon}{b_2-b_3}\\
\end{split}\label{emf}
\end{equation}

\subsection{Numerical results for $\sigma \ne 0$}
Next, we consider effects of strain inhomogeneity ($\sigma \ne 0$).   Instead of disorder averaging using the replica trick, we obtain the  phase diagram from a direct classical Monte Carlo sampling of Eq.\ref{e0}.  Working on a  40 by 40 square lattice with periodic boundary conditions, we employ a Metropolis algorithm, with an equilibration of  400000 steps, and the last 80000 steps are taken for measurement.  For each average strain $\varepsilon$, and temperature $T$, we obtain results by averaging over 14 strain configurations.  For our choice of parameters, we take $T_c=1$, $b_1=1.5$, $b_2=2/3$ and $b_3=-2/3$. We further consider $K_1/T_c=10000\gg{1}$, $K_2=K_3=K_4=K_1/3$, and $K_5=K_1/100$. In this limit, thermal fluctuations can be neglected. To avoid unnecessary finite size effects, the lattice constant needs to be rescaled with respect to $\sqrt{K_1}$. 

In our simulation, we measured the following quantities,
\begin{equation}
\begin{split}
C_{\alpha\alpha}
\equiv\overline{\frac{1}{N}\sum_i|\langle\eta_{\alpha}^*(i)\eta_{\alpha}(i')\rangle|}, \alpha=x,y.
\end{split}
\end{equation}
Here, $i$ and $i'$ denote the sites on the lattice, and $N$ is the total number of sites. $\langle...\rangle$  denotes a thermal average from Monte Carlo simulation. $i'$ is chosen for each $i$, such that the displacement between them is maximized (call it $R_{max}$). Thus, $\langle\eta_{\alpha}^*(i)\eta_{\alpha}(i')\rangle$ is the correlation function of $\eta_{\alpha}$ with the distance $R_{max}$. $C_{xx}$ and $C_{yy}$ are then the average magnitude of correlation function for $\eta_x$ and $\eta_y$ at $R_{max}$, respectively. They measure the long-range ordering of the order parameters $\eta_x$ and $\eta_y$. The phase boundary between the normal state and superconducting state ($T_c^{(1)}$) is located at $C=C_{xx}+C_{yy}=\epsilon=0.01$.  In addition, we have also studied the dependence of the correlation functions above as a function of relative position $R_{i,i'} = \vert i - i' \vert$ and have verified that the function varies exponentially with $R_{i,i'}$ for $T >T_c^{(1)}$, and is  a constant for $T < T_c^{(1)}$. 

Similarly, we define the following correlation function to characterize the breaking of  time-reversal symmetry: 
\begin{equation}
\begin{split}
C_{xy}\equiv\overline{\frac{1}{N}\sum_i\text{Re}[i\langle\eta_{x}^*(i)\eta_{y}(i')\rangle]},
\end{split}
\end{equation}
which we use  to locate the second phase boundary $T_c^{(2)}$. To do so, we add a weak symmetry breaking term $h(i\eta_{x}^*\eta_{y}+c.c.)$ with $h=-0.01$ to the free energy (note that $h$ corresponds simply to an orbital Zeeman field).

\begin{figure}[htb]
\centering
\includegraphics[width=4.2cm]{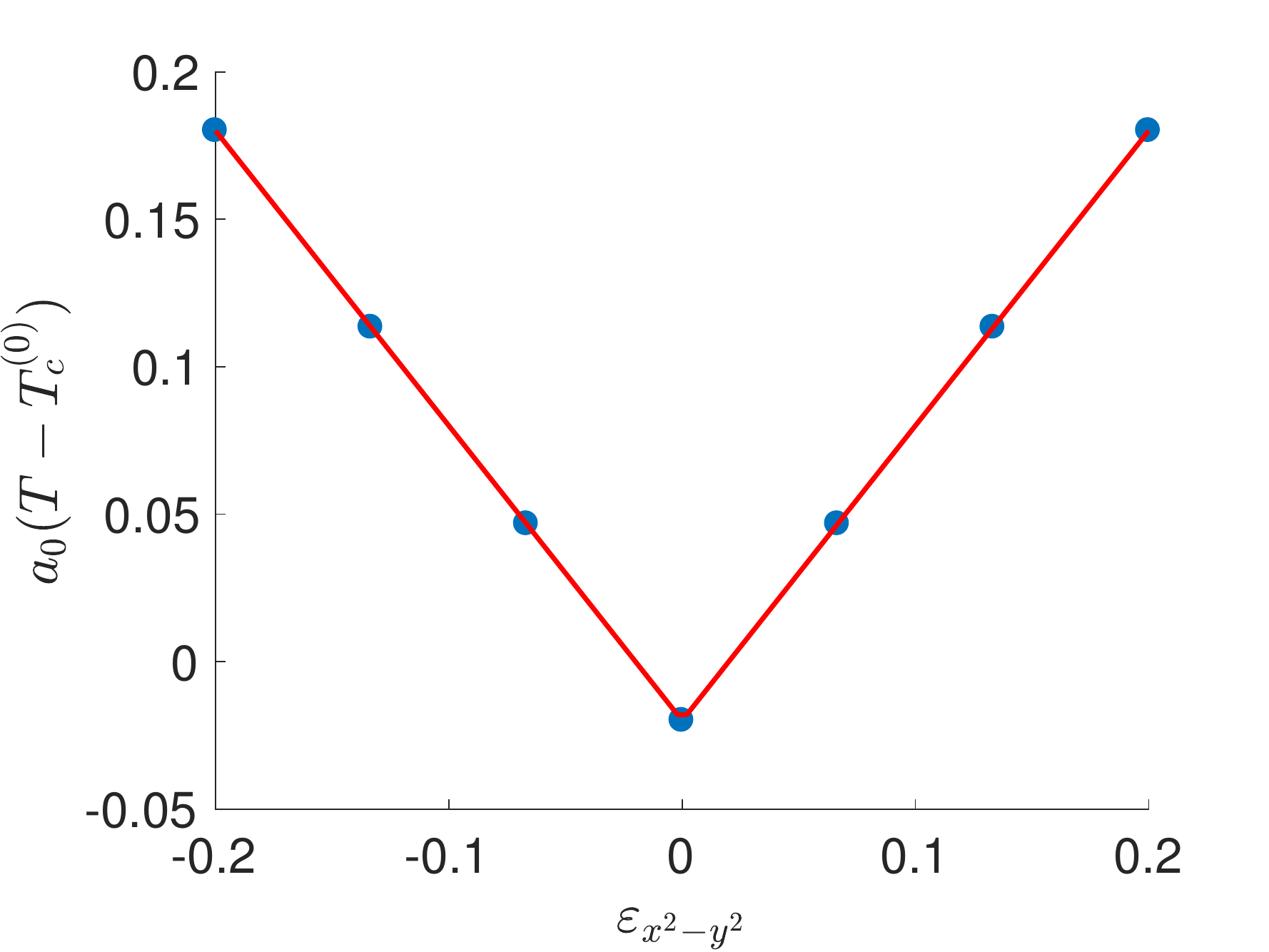}
\includegraphics[width=4.2cm]{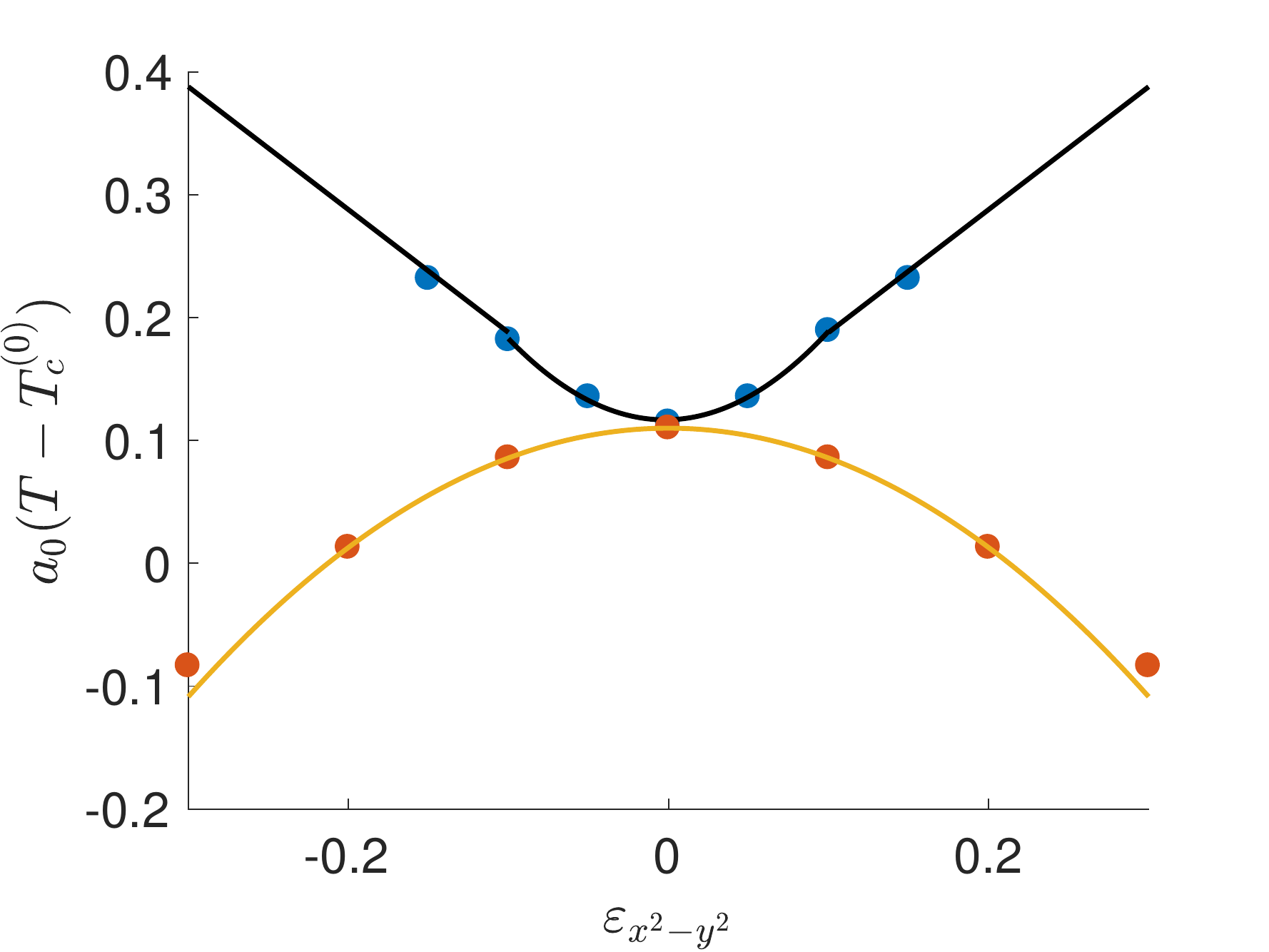}
\caption{Numerical results for the phase diagrams for (Left) $\sigma=0$ and (Right) $\sigma=0.2$, for weak thermal fluctuation. Phase boundaries $T_c^{(1)}$ and $T_c^{(2)}$ are located for 7 different average strain $\varepsilon$, using $C=C_{xx}+C_{yy}=\epsilon=0.01$, and $C_{xy}=\epsilon/2=0.005$, respectively. Solid lines are the fitted phase boundary.}
\label{f3} 
\end{figure}

The numerical results are summarized in Fig. \ref{f3}. In the absence of strain inhomogeneity, the linear phase boundary is verified. Under strain inhomogeneity, the phase boundary becomes quadratic for small $\varepsilon$. For large enough $\varepsilon$, the phase boundary stays linear.  The crossover from quadratic to linear behavior as a function of $\varepsilon$ occurs roughly at $\varepsilon_c \sim {\sigma}$.  
The reason for the quadratic response at low values of strain can be  understood as follows.  Let us consider the normal-superconducting transition.  In the case of homogeneous strain, the system is sensitive to the sign of $\varepsilon$, since the macroscopic configuration changes from a $p_x$ to a $p_y$ superconductor below the transition when the sign of $\varepsilon$ changes.  However, with strain inhomogeneities, there will always be local patches hosting both types of superconductivity.  Until the magnitude of $\varepsilon$ exceeds the rms strain set by the inhomogeneities, the macroscopic ground states are insensitive to the sign of $\varepsilon$. 
 
\section{Discussion}
\label{discussion}
In our Monte Carlo simulation, the length scale of inhomogeneity is chosen to be the lattice constant, which is much smaller than the correlation length. We focused on the correlation function at $R_{max}$, corresponding to a measurement with resolution length $R_{max}$, which is also much longer than the inhomogeneity length.
Our key result presented above is that local variations of strain have a qualitative effect on the phase diagram of a chiral p-wave superconductor.  Instead of a cusp like behavior that indicates a split transition, the transition temperature varies quadratically with strain due to such inhomogeneity.  Experimental observations of local strain variations have been reported recently in Ref. \onlinecite{Watson2018}, and constrain theories of unconventional superconductivity in this system.  

Next, we briefly comment on previous theoretical treatments.  Our study most closely resembles that of Ref. \onlinecite{Fischer2016}, where thermal fluctuations led to a quadratic variation in T$_c$ as a function of strain.  However, strain inhomogeneity was not considered, and a consequence of the theory of Ref. \onlinecite{Fischer2016}, is that at zero strain, the superconducting transition is first order.  While weakly first order transitions have been reported in an in-plane field, to our knowledge, such behavior is absent at zero field.  Furthermore, the large heat capacity jump at the transition seems to indicate to us that thermal fluctuations are relatively unimportant, and that a mean-field analysis of the superconducting transition ought to suffice.  It is precisely within such a framework that we find a quadratic variation in T$_c$ as a function of weak strain.  

\subsection{Relevance to the phenomenology of Sr$_2$RuO$_4$}
While we have shown that a chiral p-wave superconductor {\it can} exhibit a quadratic  T$_c(\varepsilon)$, it remains unclear whether Sr$_2$RuO$_4$ {\it is }indeed such a superconductor. The only experiment that directly points towards a spin-triplet (or more generally, an odd parity) superconductor is the measurement of the NMR Knight shift\cite{Ishida1998,Murakawa2004}.  Other experiments, such as the Kerr effect\cite{Xia2006} and muon spin resonance\cite{Luke1998}, require broken time-reversal symmetry, which can also arise from spin-singlet superconductivity.  These experiments therefore provide only circumstantial evidence for chiral p-wave pairing.  In addition, there are several expeimental observations that seem directly in contradiction with a simple chiral p-wave state\cite{Ishida2000,Bjornsson2005,Hassinger2017}.   Ongoing NMR studies in the presence of strain ({\it e.g.} Ref. \onlinecite{Luo2018}) are likely to shed significant light on these issues.  

  \section{Acknowledgments}
  This work was supported by the DOE Office of Basic Energy Sciences, contract DEAC02-76SF00515.  We thank  D. F. Agterberg, S. Brown, C. Hicks, M. Fischer, and R. Thomale for discussions.  
 
\bibliography{SrRuO214}{}
\bibliographystyle{utphys}

\end{document}